\documentstyle[11pt] {article}

\title{Photo-Chemical Applications of Phase-Modulus Interdependencies}
\author{
Robert Englman$^{a,b}$ and Asher Yahalom$^b$ \\
 $^a$ Department of Physics
an Applied Mathematics,\\ Soreq NRC,Yavne 81810,Israel\\ $^b$
College of Judea and Samaria, Ariel 44284, Israel\\ e-mail:
englman@vms.huji.ac.il; asya@ycariel.yosh.ac.il;}

\begin{document}
\maketitle

\newcommand{\beq} {\begin{equation}}
\newcommand{\enq} {\end{equation}}
\newcommand{\ber} {\begin {eqnarray}}
\newcommand{\enr} {\end {eqnarray}}
\newcommand{\eq} {equation}
\newcommand{\eqs} {equations }
\newcommand{\mn}  {{\mu \nu}}
\newcommand{\sn}  {{\sigma \nu}}
\newcommand{\rhm}  {{\rho \mu}}
\newcommand{\sr}  {{\sigma \rho}}
\newcommand{\bh}  {{\bar h}}
\newcommand {\er}[1] {equation (\ref{#1}) }
\newcommand{\mbf} {{ }}
\begin {abstract}
After an accolade to Joshua Jortner, we trace the influences of
his Chemistry background in his Physics writings. On the way, we
note the richness of Physics in principles (or large-scale laws)
and the fact-orientedness of Chemistry.

 Next we turn to a recent
laser-induced electron-detachment experiment and  utilize analytic
properties of the  developing wave-packet in the (complex) time
domain, studied by us previously, to relate the
 phase of the optimized laser field to its intensity. It
is suggested that these results can be used to reduce the labor of
pulse optimization in phase-intensity controlled reaction
dynamics. Phase-intensity interdependencies are also established
in simulated results (obtained with the END-algorithm) for
photo-excited hydrogen-molecules.
\end {abstract}

{Short title: Phase-Modulus Interdependencies}


\section{Joshua Jortner in Physics. An Appreciation}

Few readers will begrudge our saying some nice things about Joshua
Jortner, especially from an angle not frequently projected. Praise
has been deservedly accorded to him for the quality and importance
of his scientific works, for his leadership and for his
inspiration to his colleagues. One of us (R.E. or, in this
section, "I" for short) was some time ago (probably in the
seventies) a participant and lecturer in one of the Erice NATO
workshops. Joshua gave a course there and so did a number of other
distinguished scientists, some of whom have achieved leading
prominence on the scientific world stage. I conducted a survey
among the attendees at the Workshop, mostly postgraduate Physics
students, asking them whose course impressed them best, from whom
did they learn the most. The list of persons chosen by these young
people started with one name and ended with the same. When I
pressed them for their reason, they replied that Joshua Jortner
talked with a grand sweep, a breadth that opened for them a window
to Science, rather than just to the particular subject on which he
was lecturing.

 In continuation, I wish to delve on a "problematic" aspect of
 Joshua's works. A considerable part of these are what most of us
 would term "Physics subjects". A look at Joshua's publication's
 list reveals that these works appeared variedly in Physics
 journals, in Chemistry periodicals and in general science forums,
 which do not qualify as either. The place of the
 publication does not really matter, what is clear is  that many of his topics are
 comfortably at home in any Physics department.

 I would like to discuss this phenomenon (admittedly
 not unique to Joshua), namely that of someone with a Chemistry background
 and working in a chemist environment digging into a physics soil to
 bring up veritable treasures. Or, to use a different metaphor,
 plucking the fruits of a tree that used to be out-of-bounds, not to say forbidden, to a
 Chemist.
  Of course, it is not the ethics of this that I wish to
 discuss, but whether one can discern in these Physics papers (in their style, in the
 methodology, etc.) the upbringing of a Chemist.

 To make progress, I shall try to draw some distinctions
 between approaches that are indigenous to chemists and to
 physicists. Are there such distinctions? On the face of it, there need not be.
 After all, good science in both disciplines is expected
  to be honest, open, impersonal, self-critical and
 objective (all these things being what one likes to subsume under
  "the scientific method"). Still, there appear to be (somewhat
  unforeseen) differences, which will be seen relevant to our
  inquiry.

  According to von Neumann, $^1$ scientific laws are of
  two kinds. The first kind, called type (A), relate to {\it properties of
  the constituents} of the system under study. Thus they
  describe or predict properties (such as positions or velocities)
  of the "bodies" of which the system is composed (e.g., atoms or
  elementary articles). The second kind of laws, those of type (B), {\it characterize
   the system as
  a whole}, and do not focus on its parts. A-type laws tend to be
  specific in scope, are formal, mathematical. They are
  empirically based, but are extrapolative and flexible. They tell
  us that "one may conclude the following as well". In contrast,
  B-rules tend to be more verbal and less formal and are
  restrictive in their application. They say: "Only these are
  acceptable". (One ought to add that in the present understanding one excludes
   from "scientific laws" axioms, theorems and theories, the
   latter two being of too limited scope to qualify as laws.
   One might have called the laws "principles", it being
   understood that the laws are not only of broad scope, but
   are also widely accepted.)

  It turns out that Physics is very rich in both A- and B-rules.
  A-rules comprise Laws of Motion of bodies, the Schrodinger
  equation, the Einstein equation, the Maxwell equations,
  quarks. B-laws are causality, c-causality or subluminality,
  positivity of energy, minimal action, the exclusion principle
  (which perhaps belongs better to A), single-valuedness of
  wave-function, the correspondence principle, four laws of thermodynamics,
  Mach's principle, the principle of frame-independence, the insistence on the simplicity
  and beauty of theories and possibly others, e.g., Ref. 2.
  Of course, Chemistry has adopted many of these laws, but without
  claims of paternity. It has its own rules, too. These are, of the  A-types ,
  the laws of multiple proportions,
  the Lewis laws of structural chemistry, H\"uckel's rule, the
  Woodward-Hoffman rules. Belonging to the B-category one has the
  atomic "hypothesis" (that matter is composed of  molecules and
  of atoms) and Le Chatelier's principle. There might be only a few more
  than these.

   One sees from this, as well from the perusal of the scientific
   literature, that when one is interested in a
   broad-brushed description, the picture is that
   Physics (at least as a theoretical pursuit) is primarily principle-laden,
   whereas Chemistry is more fact-oriented. The former derives its results from a unified
   description,$^3$ the latter thrives on diversity. More
   deductive is the former ,$^4$ matched by more induction in the latter.
   (It has been said, though vigorously disputed by others such as Harry Lipkin,
   that the major advances in Physics are essentially theoretical, with experiments
   playing a confirmatory role.)

   In a recent article on the distinguished contributions to Physics by
   some former chemists and
   entitled "From chemistry to physics", L. Tisza writes "all three
   [E.P. Wigner, John v. Neumann and Edward Teller] moved from
   chemistry that seemed to them an empirical craft, toward a
   physics based on mechanics that was already penetrated by
   subtle mathematics" .$^5$ [Tisza either disregards the
   "principle-richness" of Physics or perhaps insinuates that the three
   (Jewish-descent-) Hungarians were those who were instrumental in consolidating it.]

   The above distinctions allow us to seek traces of a Chemistry background in Joshua's approach to
   physical phenomena and in his distinctive style. But before doing this, let us cite a
   historical precedent, in which Chemistry and Physics styles of writing are contrasted.
   The famous chemist, Justus von Liebig, vaunts Michael Faraday's papers, as follows:
   "I have heard mathematical physicists  deplore that the way Faraday recorded his [results]
    were  difficult to read and understood that they often
    resembled rather abstracts from a diary. But the
    fault was theirs, not Faraday's. To physicists who have
    approached physics by a road of chemistry, Faraday's memoirs
    sound like admirably beautiful music ".$^6$

    In a less serious vein, we might
    contemplate the secondary meanings frequently attached to
    "Physics" and "Chemistry". The former word has the connotation
    of being deep and fundamental, in the form of, e.g., "the
    physical reason of this is...". Recently, we have heard a
    metallurgist saying:"the physics in this is ..." when he
    clearly meant to say that "the truth in this is ....". In opposition,
    "Chemistry" has a more vital secondary usage: between two
    persons it can lead to a happy life ever after, or alternatively, can
    bring peace and understanding between warring nations.

    The "distinctive style" in Joshua's papers is unmistakable,
    especially in the Introductions and Conclusions, and it is the
    same style that impresses one in his lectures. Almost
    immediately and without warning he immerses the reader in a
    sea of facts, or in his own words, "in the rich and
    fascinating world of phenomena".$^7$ The kaleidoscope of
    phenomena usually includes as enumeration
    of instances, of effects, being part of a  broader phenomenon, of properties, of applications,
     of varieties of experimental methods and of a list of tasks
     that await to be done. And it is not just bare facts that are
     enumerated, but the characterization of each system within a
     wider context, the ordering of subjects by placing subject x smugly between subjects y
     and z. The categories are frequently labelled (A), (B), (C), etc.
     or by equivalent symbols.

     Is it unity that the author is taking his aim at? Methinks, it is rather
     the manifestation of diversification that is guiding him. He is
     less after establishing a common parentage than the
     discovering of kinships, seeking to introduce us to a big and independently functioning
     family and letting us see the role of each individual member in the group.
      If the above is correct
      (and the hypothetical factor cannot be sufficiently
      emphasized), it should leave its imprint on methodology, too.
      The deductive (and in my view, more Physics-based) approach
      normally starts with the posing and the solution of a general
      problem and takes particular sub-cases in its stride; a more
      inductive approach places the maximal effort on a specific
      instance (probably that for which data are awaiting an
      explanation) and having solved that, reminds us of further
      instances that have elements in common with the studied one.

      It is in this form that the imprints of a chemistry heritage and
environment are legible in Joshua's marvellous works in Physics.
How this comes about, has been a concern
      in the sociology of sciences. It has been proposed, that the norms of
      the reference group are accepted by the individuals forming part of
      it. In the phrasing of an eminent sociologist of science, J.
      Ben David, "[subject to some reservations] the expectations
      [in the individual] directed toward a role ... are
      determined by the institutional definition of the role".$^8$

     On more than one occasion we have heard Joshua proclaim
     the pleasure he takes at theorizing in  Physics. "The physical
     chemist dines and wines well at the table of Theoretical
     Physics." Few have repaid their meal at that table more fully
     than Joshua Jortner.

\section {Background to Reaction Control}

The enhancement of a desired end-product in a chemical (or photo-
chemical) process through controlled electromagnetic (laser)
excitation has been the aim of numerous studies. At least two
volumes comprehensively  documenting the state of the art have
recently appeared, with some articles in them that are of
particular interest to the present paper.$^{9-10}$ It emerges that
by the shaping of femtosecond laser pulses it is possible to
achieve selective molecular excitation leading to a preferred exit
channel. (A {\it caveat} on impossible pathways has also been
sounded .$^{11}$)

 A seminal idea in
the subject has been that the quantum system under investigation
could itself guide (through some automated algorithmic loop) the
search for the optimal electric field .$^{12}$ In other efforts
the relationship between the quantum state and the optimal
electric field was brought out in a different form, namely, by
explicit formulae giving the latter in terms of (overlaps of)
time-dependent wave-functions describing the evolution of the
system .$^{13,14}$
 It therefore appears from these results that some analytical properties of the molecular
 wave-function as a function of time ($t$) are carried over onto the {\it optimized}
 electric field.

On the other hand, some time ago we have investigated the analytic
properties possessed by some molecular wave-functions,
 in particular of wave-packets formed by vertical excitations between two potential
 energy surfaces, as functions of the time variable
 .$^{15-16}$
The root-cause of these analytic properties is the
lower-boundedness of the energies entering the wave-packet
:$^{17-18}$ this property induces reciprocal relations similar to
those in the Kramers-Kronig relations (which are based on
causality), but in the present case in the time domain. (The
relations are shown below in \er{RRim} and \er{RRre}).

These results are applied in the present paper to two recent works
involving photo-excitational processes, the first being
experimental, $^{19-20}$ while the second is computational, based
on the electron nuclear dynamics (END) algorithm .$^{21-22}$
However, before turning to these applications we first describe
the mathematical background of our algorithm as applied to an
optimized laser field.

\section {Reciprocity Relations for an Optimal Field}

Let a component of the electromagnetic field vector be written as
the complex quantity $E({\bf R},t)$, with ${\bf R}$ representing
position vectors\\ inside the molecule. (The field in the present
formulation  is complex: its real and imaginary parts
 are associated with the two independent polarizations of the
laser light perpendicular to the direction of its propagation
.$^9$) For all values of the position coordinates ${\bf R}$, let
$E({\bf R},t)$ satisfy as function of the complex variable t, the
conditions that it is (a) regular and (b) without zeros in the
lower half of the complex t-plane and that (c) it tends to a
constant (t-independent) finite value for large $|t|$. (The
meaning of these restrictions will be presently discussed.) Then
one has the following integral relations:
 \beq
\frac{1}{\pi} P \int_{-\infty}^{\infty}dt' \frac{\ln|E({\bf
R},t')|}{t-t'} = - \arg E({\bf R},t) \label {RRim} \enq and \beq
\frac{1}{\pi} P \int_{-\infty}^{\infty}dt' \frac{\arg E({\bf
R},t')}{t-t'} = + \ln|E({\bf R},t)| \label{RRre} \enq making the
logarithm of the field intensity and the phase Hilbert transforms
of each other. In the above equations P designates the principal
part of the integral. (For an electric field whose analyticity
domain is the upper half of the complex $t$-plane, the right hand
sides of the above equations change sign.)

To discuss experiments in which the (real) electric field is of
the form \beq E_r({\bf R},t)=F({\bf R},t)cos(\Omega t + \phi({\bf
R},t))\label{Er}\enq, where the field strength $F({\bf R},t)$ is
positive and $ \phi({\bf R},t)$ is the chirped part (in our
terminology, the analytic part) of the phase, one writes \ber
E_r({\bf R},t)
& = & Re~ F({\bf R},t)e^{i(\Omega t + \phi({\bf R},t))}\nonumber\\
& = & Re~ e^{i\Omega t} E({\bf R},t)\label {Er1}\enr where $Re$
means the real part. If the complex field $E({\bf R},t)$ thus
defined satisfies the conditions (a)-(c) above, then $F({\bf
R},t)$ and $\phi({\bf R},t))$ are Hilbert transforms, that is to
say are connected by \er {RRim}and \er{RRre}. (The alternative
definition of $E_r({\bf R},t)$, in terms of an imaginary exponent
which is the negative of that in the above equations, leads to the
same result, owing to the remark in the parentheses at the end of
the last paragraph.)

 The conditions
(a)-(c) are satisfied for a wave packet. $^{16}$ (To be accurate,
one needs to exclude from the overall phase the non-chirped part
of the wave packet, since for this the integral in \er{RRre} is
divergent and the formula meaningless; thus, this part of the
phase must be taken off the physical phase, before applying the
reciprocal equations and then reinserted in the physical
quantity). Therefore, if the optimal electromagnetic field mimics
the behavior of such wave function, it will also satisfy the
reciprocal relations. Our preliminary examination of a field that
gives the most efficient transition to an excited state wave
packet indicates that this field should indeed have both the same
modulus and the same phase as the wave-packet. (There is an
analogous result in signal theory: the {\it optimum} filter
function is proportional to the complex conjugate of the input
signal spectrum .$^{23}$)

Condition (b) is included mainly for simplicity. When it does not
hold, it can be corrected for. Furthermore when the zeros in the
lower half-plane lie far from the real t-axis, their presence is
not felt in a limited and frequently physical part of the real
time region.

Condition (c) might seem at a first sight to be un-physical, since
it requires the field amplitude to be constant (and non-zero) long
before and long after the experiment. A natural meaning of this
constant is the (temperature dependent) vacuum field intensity.
Its value nay not be readily available in many cases. This,
however, affects the results, as e.g. in the right hand side of
\er{RRim} only by the logarithm of the error committed. (For
another, somewhat similarly unexpected, appearance of quantum
electrodynamics that is needed to establish consistency, one can
call attention to the last paragraph in a landmark paper .$^{24}$)

\section {Electron Detachment in $CpMn(Co)_3$}

 A variant of control in photo-chemical processes was proposed in Ref. 25, such that
  both the intensity and the phase of the optimized laser field were adjusted.
  Carrying forward this idea, optimization of the electromagnetic field  was applied to the
  carbonyl $CpMn(Co)_3 ~(Cp=\eta^5- C_5 H_5)$, with a view of obtaining an enhancement of the
   molecular ion  product $CpMn(Co)_3^{+}$, in preference to another fragmentation
   product $CpMn(Co)_2^{+}$
.$^{19-20}$ The latter works gave accounts of a pump-probe
experiment in which the exciting field was optimized, so as to
achieve an improved ionic yield.

 The experimental arrangement in
the quoted works, described there in detail, used a laser that
sent, in the pump stage, pulses operating at 402.5nm
(24844.7cm$^{-1}$) and  of about 45 fs duration. The probe pulse
consisted of three photons of 805nm (12422.4 cm$^{-1}$) each;
these arrived with a variable delay. The optimization was effected
by a pulse shaper (operating with 256 pixels), which was capable
of simultaneous modulation of the phase (chirping) and the
intensity.

 Several potential energy
surfaces for the carbonyl molecule were shown in Fig. 3 of Ref. 19
as functions of a single nuclear coordinate (the Mn-CO distance).
In an adjacent figure (Fig.1),
\begin{figure}
\vspace{4cm} \includegraphics{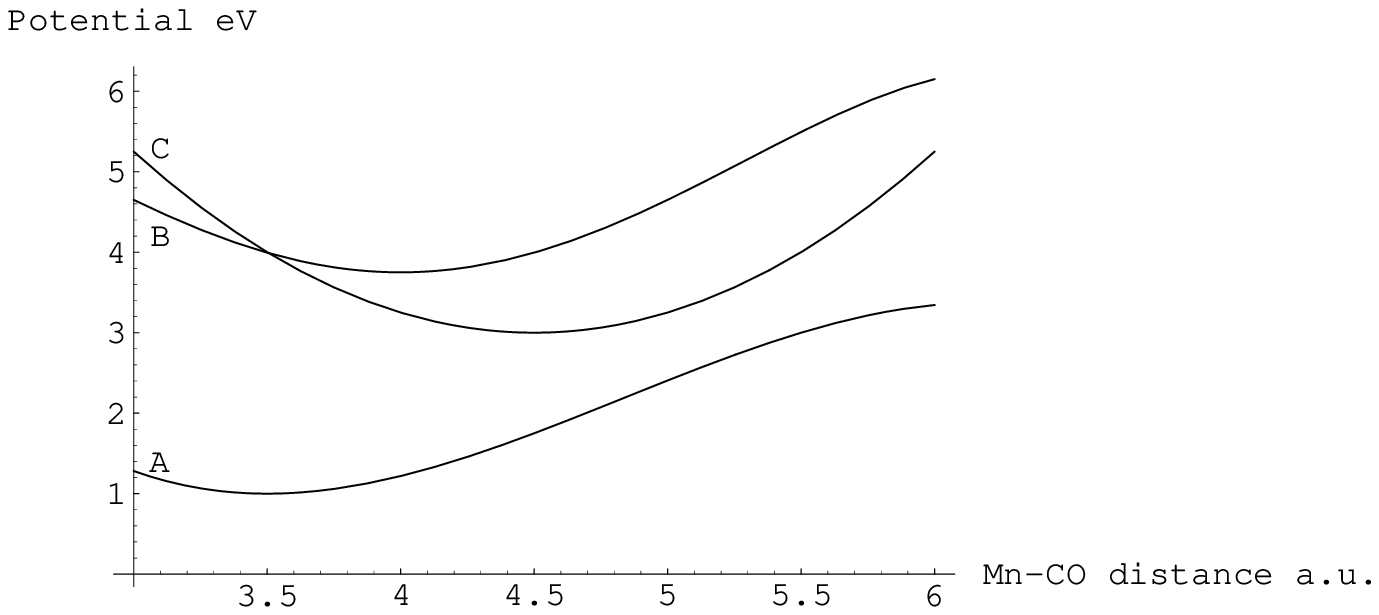} \caption { Potential surfaces (in eV)
against $Mn-CO$ distance (in a.u.) (Schematic). Potential surfaces
belong, for increasing energy heights at the left hand extremity
of the figure, to states A, B and C, respectively. The pump
excitation (in the region of 3 eV) from the ground state ($A$)
level populates levels in both excited states $B$ and $C$. In a
controlled experiment the laser field is so adjusted, as to
preferentially populate $C$.} \label {fig: nad1}
\end{figure}
we have drawn
 schematically only those three potential energy curves that are
 essential for the understanding of the
process. These are the ground state $a^1A'$, named  $A$ and the
two states in the excited state manifold, at about 3eV higher,
namely $b^1A"$ ($B$), and $c^1A"$ ($C$). It is the enhanced
population of $C$ that favors upon a further excitation (the
probe) the desired end product (the molecular ion), while
population of $B$ will promote upon further excitation of about
the same wave-length the competing process: the dissociation of
the molecule.
 Intensities and phases of the optimized electromagnetic fields are shown in the
 published curves of the above works, in particular in
Fig. 1 of Ref. 19. In addition, data values were kindly made
available to us through private communication.
 These data have enabled us to test some analytic
properties of the optimal laser field (as function of time), in
the manner explained below.

 The final conclusion from this test is
that in the optimal electromagnetic field there is a significant
(though not full) correlation between the phase and the intensity.
Theoretically, such correlation would indicate (especially, if
confirmed by more extensive tests) that the analytic properties of
the evolving molecular wave functions are imprinted on the
optimized femtosecond pulse. In this context it should be
mentioned that, though the enhancement of unimolecular reactions
by an optimal electromagnetic field has been clearly demonstrated
by now, details of its dynamics have been in need of better
understanding.$^{19}$ On a practical level, the eventual
phase-intensity correlation would mean that the optimization of
one of these does automatically lead to the optimization of the
other. Thus a considerable part of the optimization effort can be
spared.

\subsection {Optimal pump field}
The intensity and phase for the pump pulse which took place in the
interval between -100fs and about 50fs, are seen in the first half
of Fig. 1 in Ref. 19 . However, insomuch that the probe peak
follows the pump peak with a delay of about 85 fs and the width of
each excitation is also of about this value, the pump intensity
curve overlaps the probe curve. This makes the pump part of the
experimental curve difficult to consider separately. To make an
analysis of the pump curve only, we have taken from the published
figure the values of the phase between -100fs and 50fs and then
smoothly extrapolated the value of the phase outside this interval
to zero. This extrapolated pump-phase curve is shown in the
accompanying Figure 2 by a broken line.
\begin{figure}
\vspace{8cm} \includegraphics{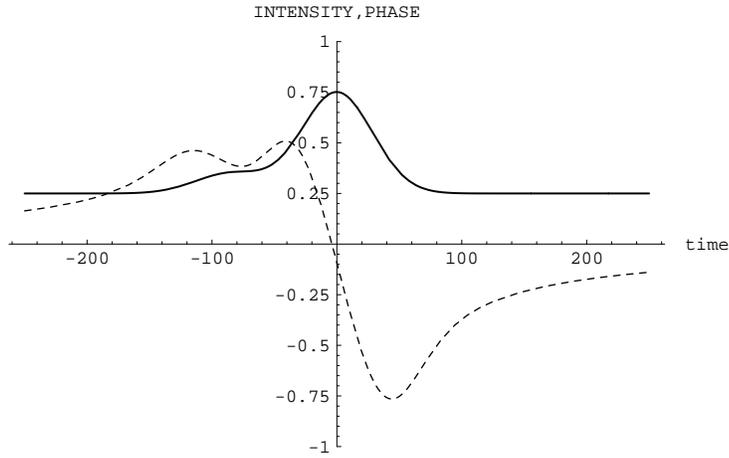} \caption { Time (in quarter-fs units) -
dependence of the pump-excitation field. Broken line: Phase (in
quarter-radians) taken from a section of Fig. 1 (in Ref. 19, but
adjusted at both extremities to a zero value. Full line: Intensity
(in arbitrary electric field strength units), calculated from the
phase through the integral relation in \er{RRre}} \label {fig: 2}
\end{figure}
In the same figure is drawn (by a full curve) the intensity of the
pump-pulse, as calculated from the second of the reciprocal
relations, \er{RRre}. The curve that this algorithm produces bears
a clear similarity to the experimental curve and indicates the
{\it ipso facto} applicability of the reciprocal-relation method.
More than this cannot be said about the method, in view of both
the theoretical and the experimental uncertainties. The main
discrepancy is in the values of the field strength in the wings:
in the experiments these tend to zero, whereas the computed
strength remains finite (representing the background). The
relatively large value is due to the rather flat phase curve. It
should be added, however, that when we attempt to apply the same
procedure [namely, calculating the intensity from the phase (or
vice versa) by the integral relations above] to a different
experiment $^{25}$,  in which the pulse was optimized for a second
harmonic generation (SHG) this leads to a complete failure, in the
sense that it comes nowhere near the optimized curves shown in
Figure 5 of the cited work .$^{25}$ That the method works with
photo-fragmentation and not with SHG reinforces our supposition
that it is the relatedness of the optimized pulse-shape to the
developing molecular wave-function that makes the method work.

\subsection {Phase-optimized pump and probe pulses}

 The data values supplied to us spanned the broad range of between
  -3000 and 3000 femto seconds and thus covered the full duration of
  pumping and "probing", as well as far beyond. The raw log-intensity
  data values
\begin{figure}
\vspace{5cm} \includegraphics{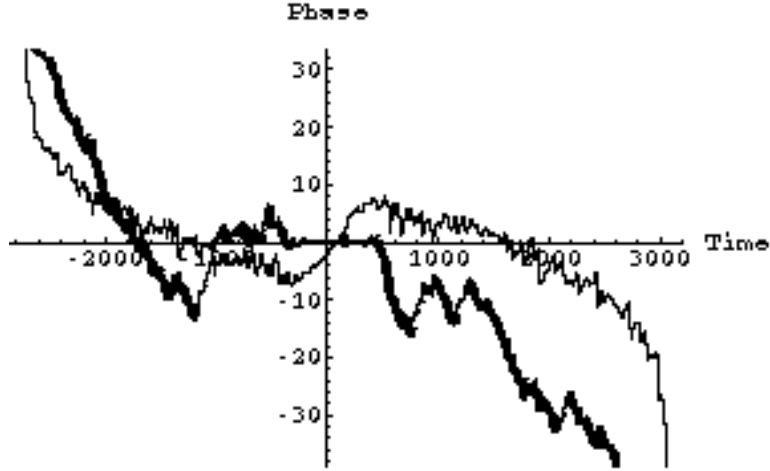} \caption { Full-scale optimal phase of
the excitation field vs time (in fs). The thin line shows the
phase as calculated from the supplied
 intensity and the formula in \er{RRim}. The thick line gives the
 phase, optimized in the experiment of Ref. 19.}
\label {fig: 3}
\end{figure}
  were inserted in the left hand side of the first of the reciprocal
   relations, \er{RRim}, and the phase calculated by integration. The
 original and calculated curves are shown in Figure 3. Again a
 significant agreement, though far from perfect in detail, is brought into
 evidence.

 The proposed interpretation of this (admittedly, partial) agreement is as follows:
 In any actual molecular wave function which arises as the solution of the
 Schr\"odinger, the phase and intensity do not vary freely, independently of
 each other. In particular, this is true of  those wave functions which
 favor electron detachment, or the
 $CpMn(Co)_3^{+}$ exit channel. On the other hand, the generation of any such wave-function
 by a external pulse of some duration, necessitates some relation between the pulse
 shape and the wave-function. The hypothesis that is being tested in this work (and
 to some extent found justified), is that it is the same integral reciprocal
 relations that hold between phase and intensity of the
 wave-function, which is also valid for the optimal field.

 \section {Excitations in Molecular Hydrogen}
 The formalism named END (Electron Nuclear Dynamics) was designed to give a
 full description of dynamics of  the electron-nuclear system dynamics
 in molecules, going beyond the Born-Oppenheimer or the adiabatic
 approximation. A review is in  Ref. 26. Of late, the
 method was applied to study the behavior of $H_3$ (as well as the
 heavier triatomic $Li_3$), when photon-energy in the range of several electron
 volts is pumped into the molecule.$^{27}$ Descriptions were
 given of the evolution of nuclear motions, which
 take place far from the equilateral triangle, equilibrium
 configuration, of the spin occupancy on each atom and of the
  amplitudes of excited electronic states (essentially of the
  $HOMO->LUMO$ excitation amplitudes), all these as functions of time.
  The time range that was investigated was typically 4 atomic units,
  or about $10^{-16}$ seconds.
  An interesting set of computational results were the
  probabilities $|c^{\alpha}(t)|^2$ and the associated phase
  shifts $\arg c^{\alpha}(t)$ (Fig. 9 in
  Ref. 26). The two quantities showed fast oscillations in
  time. What appears remarkable is that the oscillations between
  the quantities appear to be correlated, in a manner similar to that
  predicted by the reciprocal relations. However, in view of the
  complexity of the curves, as well as of the intricacy of the dynamic situation
  under study, it was thought inadvisable to try to obtain one conjugate
   quantity from the other using these curves.

 A simpler system is a rotating $H_2$ molecule, whose rotation
 speed can be regulated by varying the excitation energy of the
 molecule. Here again results were obtained by END simulations $^{27}$
 for the phase-shifts between spin-up and spin-down
 amplitudes, and also of the spin-state probabilities, when the diatomic
 was excited in the 3-4 eV range to induce a $HOMO->LUMO$
 promotion.  The time dependency of the curve was again oscillatory,
 with the maxima of magnitudes coinciding with the moments at which the
 bond was stretched .$^{26-27}$ Moreover, these results were also
  indicative of temporal
 correlations between phase and moduli of the type proposed by us.
 Still, the data were too complicated and minuscule to be
 useful to us.

 A more transparent result was obtained when the phase
 of the\\ $HOMO->LUMO$ excitation amplitude was plotted and this was compared
 with the modulus.  Figure 4
\begin{figure}
\vspace{5cm} \includegraphics{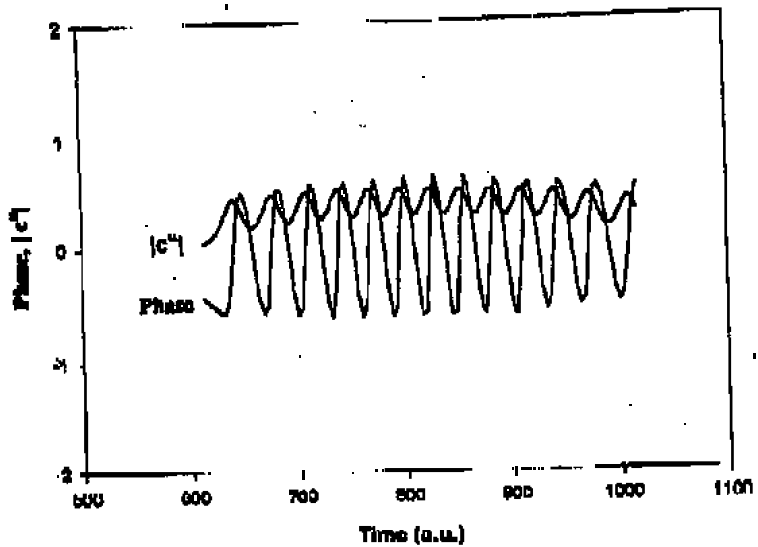} \caption { Modulus $|c^{\alpha}|$ and phase
of the $HOMO->LUMO$ transition amplitude against time for $H_2$
upon excitation with energy of 4.28 eV. Computed with END. Source:
Ref. 27.} \label {fig: 4}
\end{figure}
 shows the curves (for an excitation energy of 4.28
 eV), as given by the output. It is  seen, that the
 extrema of the moduli coincide with (numerical) maxima
 of the phase-slope, as well as  vice versa. This is a distinctive hallmark of
 the reciprocal-relations. (Cf. $\sin (t)$ and $-\cos (t)$ are Hilbert
 transforms of each other). According to F. Hagelberg,  the
 correlation can also be detected in dynamic regions where the
 phase exhibits a far less regular time dependence than in the
 time interval shown in the figure .$^{27}$

The next figure (Figure 5)
\begin{figure}
\vspace{4cm}
\includegraphics{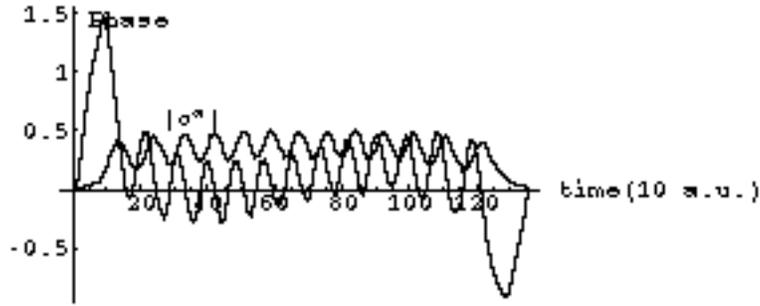}
\caption { Synthesized phase for the case of Figure 4.
The phase was calculated from the intensity using the integral
relation in \er{RRim} and the modulus values.
  The modified intensity (the input in the integral relation) also
 appears in this figure. This is the same as the $|c^{\alpha}|$
 curve shown in the previous figure, but was adjusted and extended at the wings
 to a small finite value. (Discussed in the text.)}
\label {fig: 5}
\end{figure}
illustrates the application of the reciprocal relation \er{RRim}
to this system. The "symmetric" curve is essentially the numerical
output supplied  to us for the excitation modulus, but which has
been adjusted in the wings to terminate (at both $t=\pm \infty$)
in a very small but finite value. This adjustment is necessary for
the convergence of the integrals. The results are only weakly
dependent on the value of this limit (to be precise, in a
logarithmic fashion). A physical interpretation of the limiting
value is the equilibrium value of the excited state population,
which is temperature dependent and might be extremely small, but
is nevertheless finite. The justification for this is similar to
that given at the end of section 3 for the (non-zero) asymptotic
value of the electric field intensity.

The computed output is the phase, differentiated from the modulus
by it being "anti-symmetric". This is rather similar to the
directly computed phase, which was shown in the previous figure.

However, discrepancies are evident in the wings and in the slight,
rising slope in the phase. These are connected to the
wing-adjustment in the modulus values. A final discrepancy is that
the positions of the phase-minima and maxima are interchanged
between the  directly computed (shown here in the Figure 4) and
those that are obtained from the reciprocal relation (Figure 5).
The order of the extrema in the latter with reference to the rise
or the fall of the modulus is a direct consequence of the
lower-boundedness of the energy. On the other hand, a systematic
reversal would require a "principle of upper-boundedness", which
is difficult to justify. A possible sign mistake in the END data
was investigated by their author (Frank Hagelberg) and was found
not to be the case. There remains, thus, the need for further
clarification.
\section{Conclusion}

It is very likely that in many instances in the photo-chemistry
literature interdependent behavior exists between the phase and
the modulus of some quantity. Such behavior may have been noticed
in the past, or may not. This paper examines two areas (out of
possibly many others), one in the subject of reaction-control and
the other in the dynamics of molecules, where the correlated
behavior is fairly evident (though not precise in details). We
provide a physical rationale for this correlation and give an
algorithm (based on the theory of Hilbert transforms) for its
prediction. The algorithm can have a practical use of reducing the
labor in the optimization of photo-fragmentation protocols.

\section {Acknowledgements}
Our thanks go to Frank Hagelberg, Cosmin Lupulescu and Ludger
Woeste for making their results available to us and for their
 comments, to Mark Perel'man for discussions and to Hava Nowerstern
 of the Edelstein Collection for
 help with the literature.

\end{document}